# Potential thermoelectric performance of hole-doped $Cu_2O$

Xin Chen, David Parker, Mao-Hua Du, and David J. Singh

*Materials Science and Technology Division, Oak Ridge National Laboratory, Oak Ridge, Tennessee 37831-6056, USA*

## ABSTRACT

High thermoelectric performance in oxides requires stable conductive materials that have suitable band structures. Here we show based on an analysis of the thermopower and related properties using first-principles calculations and Boltzmann transport theory that hole doped $Cu_2O$ may be such a material. We find that hole-doped $Cu_2O$ has a high thermopower of above 200 µV/K even with doping levels as high as $5.5 \times 10^{20}$ cm$^{-3}$ at 500 K, mainly attributed to the heavy valence bands of $Cu_2O$. This is reminiscent of the cobaltate family of high performance oxide thermoelectrics and implies that hole-doped $Cu_2O$ could be an excellent thermoelectric material if suitably doped.

# I. INTRODUCTION

Thermoelectrics as a field has been attracting increasing attention in recent years. Thermoelectrics are perhaps the simplest technology for direct thermal to electric energy conversion and can be used in both refrigeration and power generation devices. The current interest is stimulated in large measure by energy applications, where efficiency is a key parameter. The efficiency of thermoelectric devices is characterized by a material-dependent figure of merit $ZT = S^2\sigma T/\kappa$, where $T$ is the temperature, $S$ is the Seebeck coefficient or thermopower, $\sigma$ is the electrical conductivity, and $\kappa$ is the thermal conductivity, including both the electronic and lattice contributions. High $ZT$ is a counter-indicated property of matter, i.e. a metric that requires a combination of properties that do not normally occur together. In particular, (1) one needs a combination of high conductivity and high thermopower and (2) one needs high conductivity and low lattice thermal conductivity. It has long been understood that optimization plays a critical role. In particular, as was emphasized long ago by Ioffe,[1] the carrier concentration needs to be optimized in order to obtain the best balance between thermopower and conductivity in a given semiconductor system. With such optimization, a variety of materials with $ZT \sim 1$ have been discovered over the years. It is recognized that one way forward is through unusual electronic structures. To obtain both high conductivity and high thermopower, one seeks materials in which both light band high mobility behavior and heavy band high thermopower behavior occur at the same time. Approaches that have been proposed include dimensional reduction, as in quantum wells,[2] multiple valley band structures and band structures with sharp structure superimposed,[3] Kondo physics,[4] modification of the density of states by resonant interaction with impurities,[5] and materials with complex band structures mixing heavy and light bands[6,7] or with highly non-parabolic bands that combine heavy and light features. The state-of-the-art materials, PbTe and PbSe, may be in this latter category.[5, 8-12]

Currently, the most widely used thermoelectric materials are conventional "heavy-metal-based" semiconductor alloys, including $Bi_2Te_3$[13-15] and PbTe,[1, 5, 8-10] with $ZT$ values of 1~2. Although the high $ZT$ values in these systems make them attractive as thermoelectrics, there are potential supply issues with Te and the use of Pb is unattractive in certain applications due to toxicity. It therefore remains of interest to find non-toxic thermoelectrics based on abundant elements, provided that these new materials also have high performance. Metal oxides provide one potential avenue for this. Here we discuss p-type oxides based on monovalent Cu, using cubic $Cu_2O$ as a model.

Cu forms both divalent (e.g. CuO, La$_2$CuO$_4$) and monovalent oxides (Cu$_2$O, YCuO$_2$), with the divalent oxides being common and more stable from a chemical point of view. In the monovalent oxides, Cu has a nominally filled 3$d$ shell, which generally overlaps the top of the O 2$p$ derived valence bands in energy, although the details depend on the specific compound. The Cu 3$d$ band width is typically much narrower than the valence bands themselves. Thus one can have a situation where one has a combination of heavy primarily 3$d$ derived bands hybridized with 2$p$ character near the valence band maximum (VBM). This hybridization may be favorable for conductivity and provides a scenario that is potentially highly beneficial for thermoelectric performance. However, the fact that the divalent state of Cu is more stable than the monovalent state results in a general tendency for monovalent Cu oxides to form p-type, often in an uncontrolled way. Uncontrolled doping is disadvantageous in a thermoelectric because $ZT$ is typically a strong function of doping level. For example, delafossites such as YCuO$_{2+x}$, in spite of showing promising properties from experiment[16] and subsequent theory,[17] have not been successfully developed because they take up excess O ($x$) and become overdoped, resulting in low $ZT$.

Cu$_2$O is a well-known low-cost, nontoxic $p$-type semiconductor, also known as a mineral cuprite with a high melting point of 1508 K. It forms in a simple cubic structure, spacegroup Pn-3m. Its optical and semiconducting properties have attracted much attention. However, there is little reported on the thermoelectric properties of hole-doped Cu$_2$O, although the oxide of the neighboring element, ZnO, has received substantial interest as an n-type thermoelectric at high temperature.[18-24] Importantly, Cu$_2$O is an example of a monovalent Cu oxide that, while invariably p-type, forms with low carrier concentrations. As such it may be amenable to controlled doping to higher, but still limited, carrier concentrations appropriate for thermoelectric application.

Here we use a combination of first-principles calculations and Boltzmann transport theory[25-27] to investigate the thermoelectric potential of hole-doped Cu$_2$O. We find that the thermopower of this material can reach above 200 µV/K while at the same time having doping levels as high as $5.5\times10^{20}$ cm$^{-3}$ at 500 K. This behavior can be mainly attributed to the heavy valence bands in Cu$_2$O. These results suggest experimental investigation of methods for obtaining these doping levels in Cu$_2$O and measurements of thermoelectric properties.

## II. APPROACH AND RESULTS

The electronic structure calculations were performed using the full-potential linearized augmented plane-wave (LAPW) method,[28] as implemented in the WIEN2K code.[29] We employed the experimental lattice constant of 4.27 Å, and the radii of Cu and O LAPW spheres were chosen to be 1.84 and 1.63 bohr, respectively. We did convergence tests for the basis set and Brillouin zone sampling. Based on these we used $RK_{max}$= 9 ($R$ is the smallest LAPW sphere radius, i.e. 1.63 bohr and $K_{max}$ is the cut-off for the interstitial planewave sector of the basis) with a 12×12×12 **k**-point mesh for the self-consistent electronic structure calculations (a denser final mesh was used for the transport calculations, as discussed below). We used the Perdew-Burke-Ernzerhof (PBE)[30] generalized gradient approximation (GGA). Figures 1 and 2 show the calculated band structure and electronic density of states (DOS), respectively. These results are in good accord with those obtained previously with this method.[31-34] As shown in Fig. 1, $Cu_2O$ has a direct band gap of ~0.6 eV at Γ point with this approximation for the exchange correlation function. Our calculated band gap is in good agreement with other theoretical results using standard density functional methods,[34, 35] but it is much smaller than the experimental gap of 2.17 eV,[36] as expected since density functional methods usually underestimate semiconductor band gaps. $Cu_2O$ is a material where this problem is particularly large. However, $Cu_2O$ is a well-studied material, and a key point that emerges from past work is that the shapes of valence bands near the VBM obtained with different functionals are very similar although the band gaps of course differ.[35, 37] For thermoelectric properties, the crucial part is within a few (~5-10) $kT$ of the band edge, where $T$ is the temperature. This can be simply rationalized in terms of the electronic structure (see below). Convergence of the Brillouin zone sampling is important for obtaining reliable thermoelectric properties. Therefore, for the transport calculations we employed the electronic structure derived from PBE-GGA, except that the band gap was corrected to the experimental value using a scissors operator to prevent bipolar transport at high temperature. Calculated valence band masses along the Γ-X direction for the top three bands are 3.5, 3.5, and 0.2 $m_0$, where $m_0$ is the free electron mass. As is well known, such heavy masses are highly favorable for the thermopower and thermoelectric performance,[38] and a mixture of heavy and light bands is also favorable.[6, 39]

As shown in Fig. 2, the electronic density of states (DOS) in the energy range between ~-8 to ~-5 eV, is derived from O 2$p$ electrons, while the top of the valence bands from ~-4 eV to the

VBM, are primarily from Cu 3$d$ states hybridized with O 2$p$ states. This provides an explanation for the relative similarity in the upper valence bands as found in prior work with different treatments.[35, 37] In Cu$_2$O the main Cu 3$d$ bands are above the O 2$p$ as is seen in the projected density of states. The main source of error in the band gap of $d^{10}$ compounds such as Cu$_2$O and ZnO is an incomplete cancelation of the self-interaction of the $d$-orbitals. Thus the $d$ orbitals are placed too high in energy. A better treatment will increase their binding energy. In the case of Cu$_2$O this will lower the $d$ bands towards the O $p$ bands, while in ZnO that $d$ bands move lower below the O $p$ bands. In Cu$_2$O this lowering will not be enough to change the character of the upper valence bands since the band gap correction is smaller than the separation between the O $p$ and Cu $d$ manifolds. Furthermore, one may note that there will be a partial cancelation of two effects on the hybridization as one improves the treatment of correlation – (1) the above mentioned increased 3$d$ binding energy (and band gap), bringing the $d$ and $p$ bands closer together and tending to increase hybridization and (2) increased localization of the $d$ shell, tending to reduce hybridization.

To better elucidate this, we have also performed DOS calculations within a hybrid functional approach, using the Heyd-Scuseria-Ernzerhof (HSE) functional[40] as implemented in the VASP code[41, 42] and compare with the PBE results and experimental photoemission data. Figure 3 shows the calculated results for the valence band edge region that is important for transport (note that thermoelectric properties will depend on the electronic structure within ~5-10 $kT$ of the VBM) along with the Cu$_2$O photoemission data of Shen and co-workers,[43] which agrees with other photoemission studies of Cu$_2$O.[44-46] Note that Shen and co-workers state an experimental resolution for their ARPES experiment of 0.25 eV and we have therefore incorporated a Gaussian broadening, using this 0.25 eV value, into both density-of-states plots. For this reason the calculated broadened DOS do not vanish at the energy of the VBM.

As shown in Fig. 3, both PBE and HSE DOS agree remarkably well with the ARPES intensity data in the first 0.5 eV below the VBM, which is the relevant region for p-type transport in the doping and temperature range considered here. Furthermore, angle resolved photoemission[46] show both the heavy bands and a light band at the band edge as predicted. Such good agreement is noteworthy considering that we have not made any adjustments of the experimental data, and strongly suggests that the calculated PBE GGA transport results will be reliable.

Returning to the calculated PBE band structure, following ligand field theory, the Cu $d$ derived bands are formally anti-bonding combinations of Cu $d$ and O $p$ states, with the strongest anti-bonding character anticipated at the valence band edge. This is seen in the partial O 2*p* character of the DOS, which is larger between the -2 eV and 0 eV relative to the band edge than at higher binding energy. As mentioned, such mixing is generally favorable for conductivity. In any case, the transport properties of hole-doped $Cu_2O$ are closely related to the mixture of Cu 3*d* bands hybridized with O near the VBM. As is depicted in Fig. 1, the highest valence band shows a small band width of less than 0.75 eV. As mentioned, this is in accord with photoemission experiments, which show a peak with the corresponding width at the valence band edge. This leads to a high DOS near the band edge. Such low dispersion bands are very favorable for thermoelectric performance provided that the material can be doped and is electrically conductive. This is qualitatively similar to the situation in the high *ZT* oxide thermoelectric $Na_xCoO_2$,[47] although in addition to narrow bands derived from hybridized 3*d* and 2*p* orbitals, that system is also near magnetism.[48] Returning to $Cu_2O$, the DOS in the valence bands has a prominent peak at approximately 0.5 eV below the VBM. This can be attributed to the remarkably flat band along the Δ, Z and Σ directions. In fact, such heavy mass bands and high DOS near the Fermi level strongly imply large thermopower of hole-doped $Cu_2O$. Also, we note that such a sharp maximum in the DOS near the band edge has been discussed as an "ideal" electronic structure for a thermoelectric,[3] although considering the likely temperature range for application of $Cu_2O$ it would be better if the peak were closer to the band edge.

We calculated transport properties within Boltzmann transport theory. We did this based on the converged electronic structures, using the BoltzTraP[49] code. For this purpose the electronic states were calculated on a much more dense 34×34×34 **k**-point mesh in Brillouin zone. The use of dense grids is important for obtaining reliable transport properties. The constant relaxation time approximation was used in these calculations. With this approximation the thermopower can be determined as a function of doping and temperature without any adjustable parameters. This approximation, which consists of assuming that energy dependence of the scattering rate can be neglected at fixed doping and temperature, has been successfully used to provide a good description of transport properties in a variety of thermoelectric materials.[6, 11, 12, 27, 50-55] We note that it does not involve any assumption about the dependence of the scattering rate on either temperature or doping level.

We present the calculated thermopower as a function of doping at various temperatures from 200 K to 900 K in Fig. 4. We note that the thermopower increases with temperature over the whole range even at low doping levels and high temperature, indicating that there is no bipolar conduction as expected from the substantial band gap. It is clear from Fig. 4 that high values of thermopower ($S > 200$ µV/K) are obtained even at large carrier concentrations of $1.3 \times 10^{21}$ cm$^{-3}$ in a wide temperature range up to 900 K. For comparison, typical semiconductor thermoelectrics such as PbTe are optimized at order of magnitude lower carrier concentrations. In this respect, Cu$_2$O is more similar to Na$_x$CoO$_2$. The high thermopowers at these doping levels derive from the heavier Cu-3$d$ band and larger DOS in the valence band of Cu$_2$O, as expected from the above discussion.

We note that Cu$_2$O can decompose in air at elevated temperatures.[56] While this decomposition reaction is also dependent on oxygen pressure,[57] and perhaps could be mitigated in applications, we focus on the lower temperature range below 500 K where the decomposition of Cu$_2$O could be neglected.[56] According to our calculations, at 500 K thermopowers in excess of 200 µV/K are obtained up to $p = 5.5 \times 10^{20}$ cm$^{-3}$. Even at the lower temperature of 300 K, this thermopower can be achieved at very heavy hole concentration of $2.5 \times 10^{20}$ cm$^{-3}$. The current results show that Cu$_2$O with heavy hole-doping has a good thermoelectric potential in a wide temperature range. Obviously, since lattice thermal conductivity decreases with $T$, while because of the large band gap, the thermopower and almost certainly the power factor $\sigma S^2$ increases with $T$, the best $ZT$ will be at high temperature, i.e. 500 K, or higher if the material is stabilized at high $T$. This is because in the usual regime for heavily degenerate doped materials at moderate to high temperature (lattice thermal conductivity controlled by anharmonic Umklapp phonon scattering yielding $\kappa_l \sim 1/T$, conductivity controlled by electron phonon scattering, $\sigma \sim 1/T$, and electronic thermal conductivity following Wiedemann-Franz, $\kappa_e = L_0 \sigma T$, $\kappa = \kappa_e + \kappa_l$), with $S$ increasing as calculated here, $ZT$ will be a strongly increasing function of $T$. We note that since the decomposition of Cu$_2$O in oxygen environments is by oxidation (related to Cu$^+$ → Cu$^{2+}$), it may well be that heavy p-type doping, which is essentially oxidation, will stabilize Cu$_2$O to higher $T$.

While the thermopower can be obtained without adjustable parameters using the constant relaxation time approximation, this is not the case for the electrical conductivity. The electrical conductivity intrinsically involves a scattering time $\tau$, through the expression for the transport function $\sigma(E) = N(E)v^2(E)\tau(E)$, where N is the density of states and v the Fermi velocity, and the corresponding expression for the electrical conductivity $\sigma(T) = \int \sigma(E) df/dE$, where f is the Fermi function. Unfortunately, our first principles calculations do not give the actual scattering time, so we need to treat this as a parameter. However, this scattering time is not likely to depend on energy as strongly as $N(E)v^2(E)$ (which for a parabolic band is proportional to $E^{3/2}$, where E is measured from the band edge) or *df/dE,* which is a sharply peaked function around the chemical potential. Hence, as a first approximation for optimizing the actual power factor $S^2(T)\sigma(T)$, we here optimize the quantity $S^2(T)\sigma(T)/\tau$. Given the general lack of experimental information regarding the doping dependence of the hole mobility of $Cu_2O$, further approximations concerning $\tau$ are not likely to yield significant additional insight. We do make an effort to account for the change in $\tau$ (or, equivalently carrier mobility) with doping in the next section.

In order to present more information on thermoelectric behavior, we calculate the electrical conductivity and the power factor divided by the inverse scattering rate, *τ* at 300 K and 500 K, as illustrated in Figs. 5 and 6, respectively. It can be clearly seen from Fig. 5 that the optimal doping level at 300 K corresponding to the peak power factor is $5.7\times10^{20}$ cm$^{-3}$, equivalent to ~0.04 holes per unit cell if $\tau$ would be independent of doping level. At this doping level, the calculated $\sigma/\tau$ is around $10^{19}$ $(\Omega ms)^{-1}$. At 500 K, the maximum power factor of $Cu_2O$, neglecting the doping dependence of $\tau$, would occur at a heavy doping of $1.2\times10^{21}$ cm$^{-3}$, or ~0.09 holes per unit cell. More likely, $\tau$ decreases with increasing doping level, which would shift the peak power factor in the direction of lower doping.

## III. DISCUSSION

The current results discussed above suggest that heavily hole-doped $Cu_2O$ possesses high thermopowers at a wide temperature range, suggesting good thermoelectric performance if appropriately doped. Compared to its neighboring compound ZnO,[22] the well-known oxide thermoelectric, $Cu_2O$ has much lower experimental lattice thermal conductivity and a much more favorable electronic structure. However, to achieve a high *ZT* value for practical applications, the mobility of $Cu_2O$ needs to be high enough to ensure a good electrical conductivity. Whether this is the case is not known. Quantitatively, if a mobility of 95 cm$^2$/Vs were achieved at the heavy doping of $2.5\times10^{20}$ cm$^{-3}$, an electrical conductivity of $3.8\times10^5$ $(\Omega m)^{-1}$ would result. High mobility near 100 cm$^2$/Vs at room temperature have been reported on $Cu_2O$ nanowires (> 95 cm$^2$/Vs),[58] polycrystalline thin films,[59, 60] and single crystals,[61] but these are at a doping level far from those discussed here, and so can only be taken as indicating a possibility of high mobility. In combination with lattice thermal conductivity of 4.5 W/mK[62] and our predicted room-temperature thermopower of 200 µV/K at this doping level, we can obtain a high *ZT* value of ~0.6. Considering the 1/T dependence of $\kappa_l$ and mobility, at 500 K $\kappa_l$ may be reduced to 2.7 W/mK, and the mobility should be 57 cm$^2$/Vs according to the above assumed room-temperature mobility of 95 cm$^2$/Vs. Then, a higher *ZT* value of 1.1 could be obtained at 500 K at a heavy doping of $5.5 \times 10^{20}$ cm$^{-3}$.

As mentioned, the actual value of the mobility at the heavy dopings envisioned here is a significant source of uncertainty regarding the ZT estimates, which we now discuss. In general, carrier mobility decreases with increasing carrier concentration for degenerately doped semiconductors (see, for example, Fig. 7 of Ref. 63) and then reaches a limiting value in the non-degenerate limit. The non-degenerate limit is approached as the temperature-dependent chemical potential moves into the band gap and is typically quantified by a parameter $\eta = \mu/T$, where $\mu$ is the chemical potential measured from the band edge; we follow the usual convention[64] that $\eta$ is negative when the chemical potential lies within the gap. Typically the degenerate limit is approached for $\eta > 2$.

However, for the doping level ($5.5 \times 10^{20}$ cm$^{-3}$) and temperature (500 K) specified above, Cu$_2$O is rather far from the fully degenerate limit – $\eta$ takes the intermediate value 0.3 (see Fig. 7), which is neither fully degenerate nor non-degenerate. Hence it is unclear exactly how much of a carrier mobility reduction, relative to the non-degenerate regime, would occur.

To put this quantitatively, let us assume a 75 percent reduction in the mobility relative to the nanowire case, and retain the assumed 1/T dependence of the mobility. Then the 1.1 500 K ZT figure mentioned above becomes 0.6, based on the Wiedemann-Franz relation, outlined below. (The reason this ZT figure is not one-fourth of 1.1 is that there is a concomitant 50 percent reduction in the thermal conductivity as a result of the electrical conductivity reduction, which raises ZT significantly.) While this may appear unimpressive, it is in fact the ZT of the commercially used, doping-optimized Bi$_2$Te$_3$,[65] without alloying or bulk nanostructure optimization. Alloying of Bi$_2$Te$_3$ with Se or Sb raises ZT to unity, by lowering the lattice thermal conductivity, and when bulk-nanostructured[13, 14, 66] to 1.5 – a value two and a half times larger than the merely doping-optimized value. A similar improvement should be possible for Cu$_2$O, which would place it among the highest performing materials in the 500 K temperature range. Hence it is plausible that full optimization of Cu$_2$O could yield a high performance thermoelectric.

It should be clear from the above discussion that a definitive assessment of carrier mobility at the heavy dopings envisioned here can only be made via experimental measurements, which have not yet been done. A mobility reduction at these dopings, relative to the non-degenerate case, of as much as 90 percent could take place and would significantly reduce ZT from the above estimate if present. It is also true, however, that at 500 K ionized impurity scattering, which typically has a weaker temperature dependence than 1/T, could be a significant contributor, implying a higher mobility at elevated temperature than if only electron-phonon scattering were present. Ultimately experiments on appropriately doped samples can answer this question and we necessarily leave it to future study.

Regarding the thermal conductivity, we note that the lattice thermal conductivity of $Cu_2O$ at 300 K of 4.5 W/mK is high for a thermoelectric, especially compared with the well-known heavy-metal-based thermoelectric materials such as PbTe. In the expression of *ZT*, if we substitute $\sigma$ by electronic thermal conductivity $\kappa_e$, given by the Wiedemann-Franz relation: $\kappa_e = L_0\sigma T$, then *ZT* can be written as $ZT = (\kappa_e/\kappa)S^2/L_0$, where $L_0 = (156\ \mu V/K)^2$, is the Lorenz number. Obviously, if the lattice thermal conductivity can be reduced without a corresponding reduction of the electronic term, higher *ZT* values will result (the limit with *S*=200 μV/K is *ZT*~1.6). It is important to note that $Cu_2O$ is in a different regime than e.g. PbTe, i.e. an order of magnitude higher doping level. In this limit, the electrical conductivity may be higher, which would set a higher scale for the lattice thermal conductivity.

## IV. CONCLUSION

In summary, a detailed study of the transport properties for hole-doped $Cu_2O$ is carried out using the full-potential LAPW method and Boltzmann transport theory. The results show that hole-doped $Cu_2O$ has a high thermopower at heavy doping levels, suggesting that it may be a high *ZT* thermoelectric material if it can be heavily doped to achieve good conduction. Considering the simple cubic structure of $Cu_2O$, and known chemical properties of $Cu_2O$ it may be possible to dope the compound to suitable carrier concentrations. We hope that our current findings will stimulate future experimental exploration of the thermoelectric properties of $Cu_2O$ and other oxides containing monovalent Cu.

# ACKNOWLEDGMENTS

This work was supported by the Department of Energy, Office of Science through the S3TEC Energy Frontier Research Center.

# Figure captions

**FIG. 1.** Calculated band structure of $Cu_2O$ as given with PBE-GGA. Zero energy is set at the valence band maximum.

**FIG. 2**. (Color online) Calculated electronic density of states and projections onto the Cu-*d* and O-*p* spheres.

**FIG. 3**. (Color online) Comparison of calculated densities-of-states using the PBE and HSE functionals with the ARPES photoelectron intensity of Shen et al. Note that an 0.25 eV broadening has been applied to the densities-of-states, consistent with the stated experimental resolution of Shen et al.

**FIG. 4**. (Color online) Calculated thermopowers for hole-doped $Cu_2O$ as a function of carrier concentrations at various temperatures.

**FIG. 5**. (Color online) Calculated electrical conductivity and power factor results with respect to $\tau$ at 300 K.

**FIG. 6**. (Color online) Calculated electrical conductivity and power factor results with respect to $\tau$ at 500 K.

**FIG. 7**. (Color online) Calculated reduced Fermi level vs. hole concentration at 500 K.

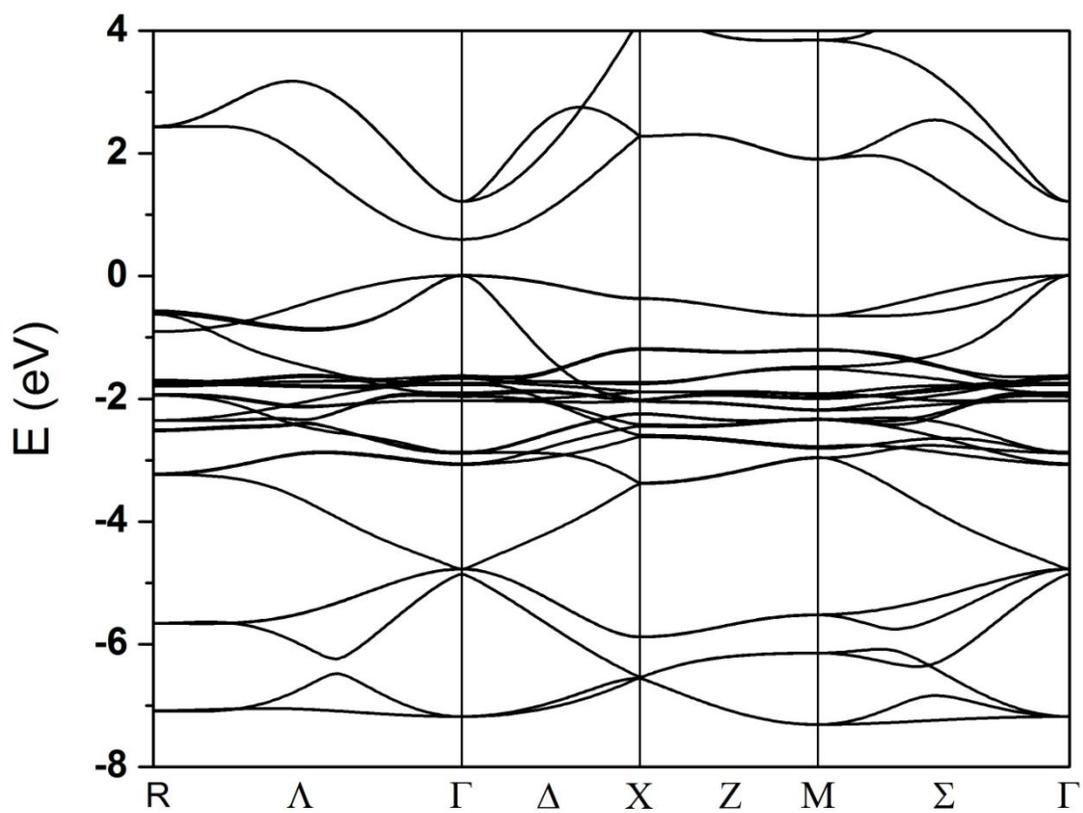

**Fig. 1**

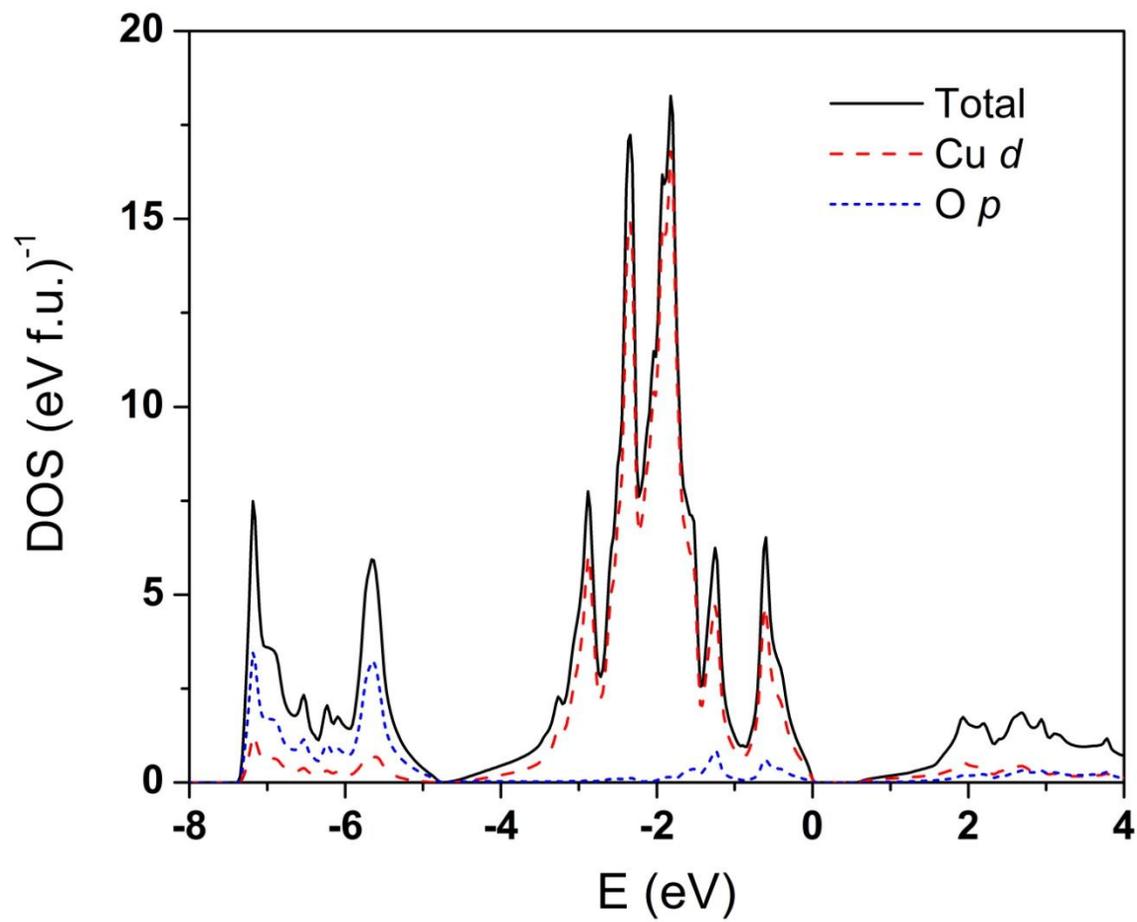

**Fig. 2**

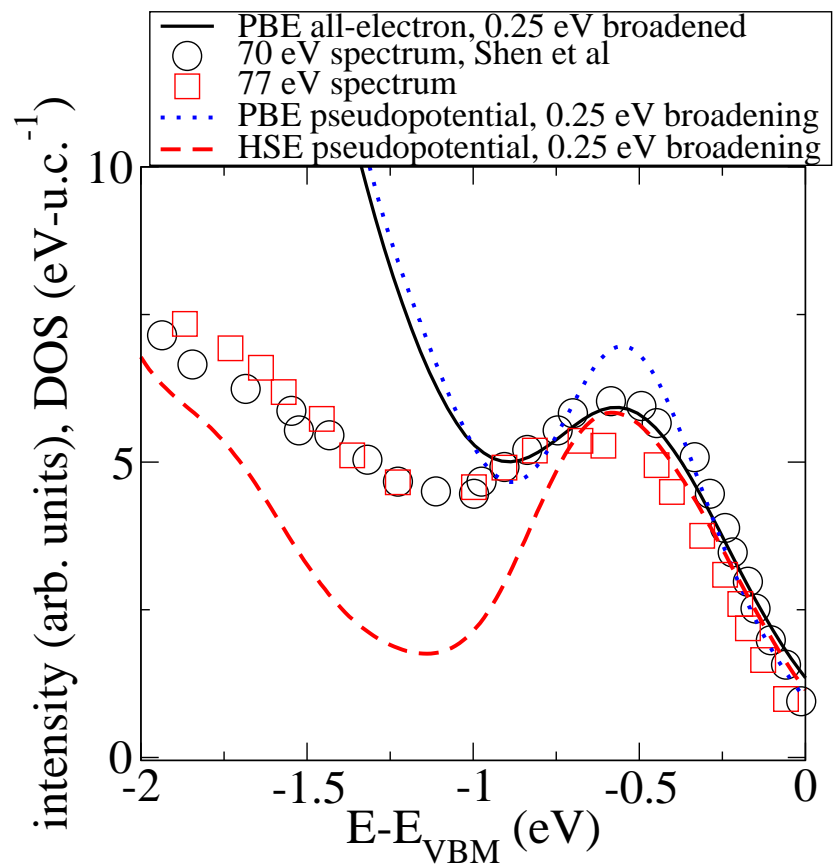

**Fig. 3**

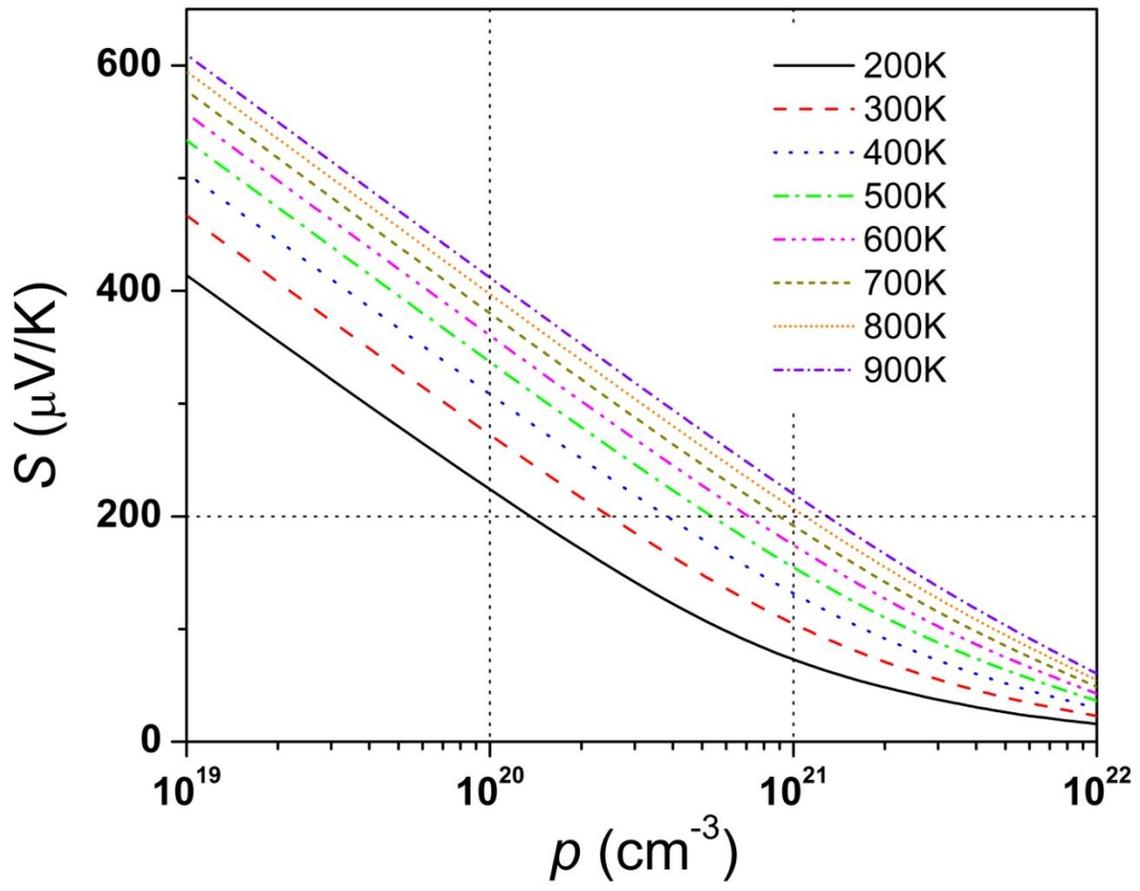

**Fig. 4**

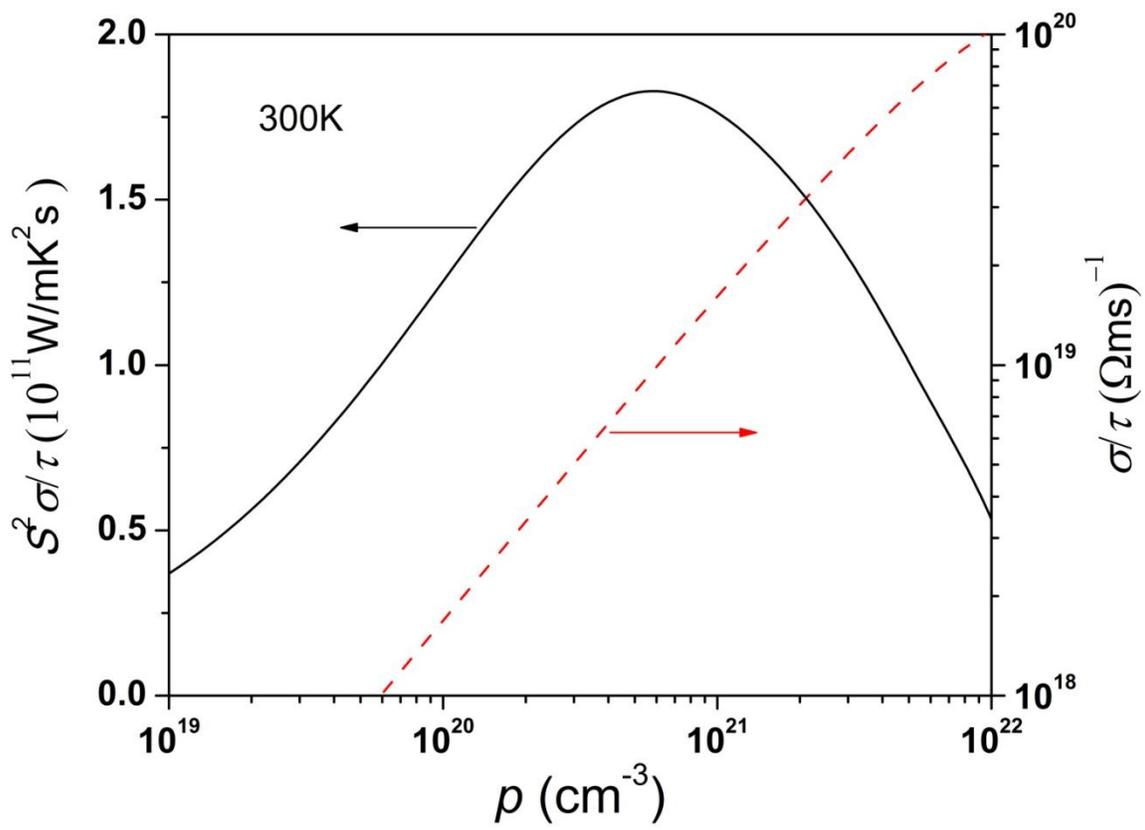

**Fig. 5**

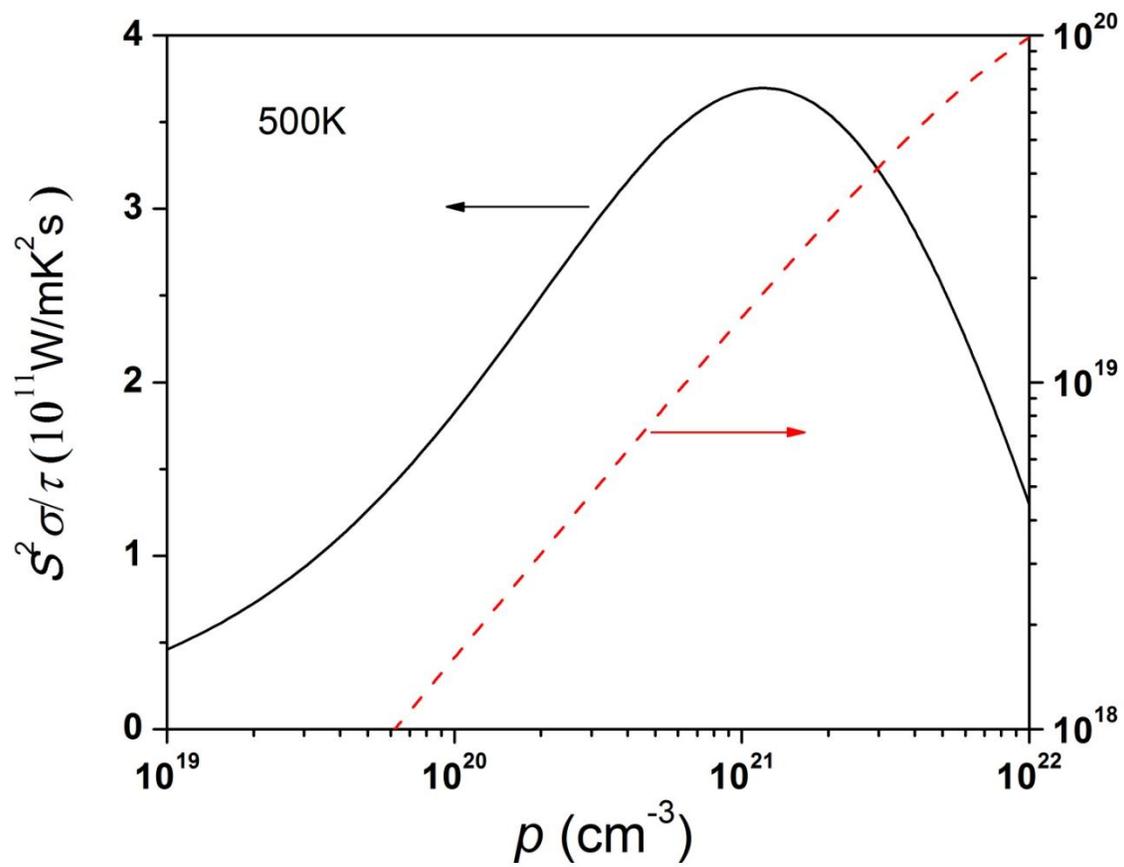

Fig. 6

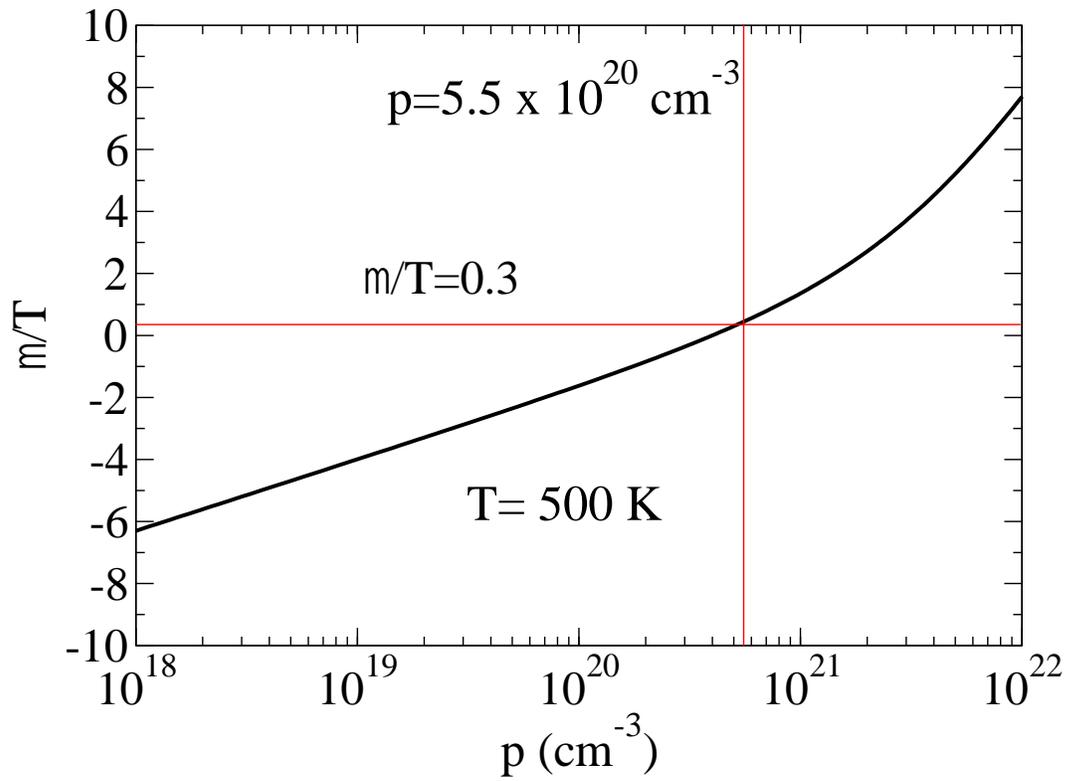

**Fig. 7**